\documentclass[document]{aa}

\usepackage{amsmath,amsfonts,graphicx,amssymb}
\usepackage{listings,rotating,float}
\usepackage{xcolor, textcomp}
\usepackage{natbib}
\usepackage[nottoc]{tocbibind}
\usepackage{enumitem}
\usepackage[nottoc]{tocbibind}
\usepackage{savesym}
\savesymbol{tablenum}
\usepackage{siunitx}
\restoresymbol{SIX}{tablenum}
\usepackage{bm}
\usepackage{aas_macros}

\definecolor{mygreen}{RGB}{28, 172, 0}
\definecolor{mylilas}{RGB}{170, 55, 241}

\numberwithin{equation}{section}

\begin{document}

\title{The need for active region disconnection in 3D kinematic dynamo simulations}

\author{T. Whitbread\inst{1}
	\and
	A.~R. Yeates\inst{1}
	\and
	A. Mu\~noz-Jaramillo\inst{2}\fnmsep\inst{3}\fnmsep\inst{4}
}

\institute{Department of Mathematical Sciences, Durham University, Durham, DH1 3LE, UK\\
	\email{anthony.yeates@durham.ac.uk}
	\and
	Southwest Research Institute, 1050 Walnut St. \#300, Boulder, CO 80302, USA\\
	\email{amunozj@boulder.swri.edu}
	\and
	National Solar Observatory, 3665 Discovery Drive, Boulder, CO 80303, USA
	\and
	High Altitude Observatory, National Center for Atmospheric Research, 3080 Center Green, Boulder, CO 80301, USA}         

\date{Received ; Accepted }

\abstract{In this paper we address a discrepancy between the surface flux evolution in a 3D kinematic dynamo model and a 2D surface flux transport model that has been closely calibrated to the real Sun. We demonstrate that the difference is due to the connectivity of active regions to the toroidal field at the base of the convection zone, which is not accounted for in the surface-only model. Initially, we consider the decay of a single active region, firstly in a simplified Cartesian 2D model and subsequently the full 3D model. By varying the turbulent diffusivity profile in the convection zone, we find that increasing the diffusivity -- so that active regions are more rapidly disconnected from the base of the convection zone -- improves the evolution of the surface field. However, if we simulate a full solar cycle, we find that the dynamo is unable to sustain itself under such an enhanced diffusivity. This suggests that in order to accurately model the solar cycle, we must find an alternative way to disconnect emerging active regions, whilst conserving magnetic flux.}

\keywords{Diffusion -- Dynamo -- Magnetohydrodynamics (MHD) -- Sun}

\maketitle

\section{Introduction} \label{intro}

A major goal of solar physics is to understand the solar cycle: the near-periodic rise and decline of solar magnetic activity. Periods of maximum activity correspond to more frequent space weather events like solar flares and coronal mass ejections,  which pose threats to satellites and astronauts and can cause technological disruption at Earth. On the other hand, the state of the Sun's magnetic field at solar minimum gives us an indication of the general global behaviour of the next solar cycle \citep{schatten78,andres13,pesnell16}. It is generally accepted that solar magnetic activity is maintained by a dynamo mechanism. However, it is not yet possible to directly measure the magnetic fields in the solar interior. Instead we currently rely on mathematical models to provide insight into the dynamo process.

There are numerous varieties of dynamo model, each having their own strengths and limitations (for reviews see \citealp{charreview2,charreview}). Here, however, we focus on Babcock-Leighton (B-L) models \citep{babcock,leighton64,leighton69}. In the B-L regime, toroidal field is converted to poloidal field via the emergence and decay of active regions at the photosphere. The cross-equatorial cancellation of leading polarity flux and subsequent preferential polewards transport of trailing flux results in a polarity reversal of the polar field. The polar field is then pumped down into the convection zone, where it is sheared back into toroidal field by differential rotation and transported equatorwards by the returning branch of meridional circulation or latitudinal turbulent magnetic pumping, becoming the seed field for the next cycle. An appealing property of B-L dynamos is that they operate in line with observations of the photospheric radial magnetic field \citep{wangetal89b,wangsh91}. Furthermore, they have been found to reproduce features of the solar cycle \citep{dikpati04,mackayreview}.

Progression in this area thus far has primarily been through the implementation of 2D or 2$\times$2D B-L dynamo models \citep[e.g.][]{wangetal91,durney95,chatterjee04,guerrero08,lemerle2,bhowmik18}. However, we would ideally like to develop 3D B-L dynamo models in order to realistically model the emergence of buoyant magnetic structures and fully describe the evolution of magnetic fields under the effects of diffusion, differential rotation, and meridional circulation. These models are more complex and require in-depth calibration in order to match the observed magnetic field. Nevertheless, success in overcoming these obstacles would be a sizeable step towards the development of a forecasting model for the Sun-Earth system \citep{nitaetal}. This would hopefully provide us with the most accurate solar cycle predictions to date.

\citet{kd3} developed KD3, a 3D kinematic B-L dynamo model. In KD3, the MHD induction equation describes the evolution of the mean magnetic field:
\begin{equation} \label{induction2}
  \frac{\partial{\mathbf{B}}}{\partial{t}} = \nabla \times \left(\mathbf{v} \times \mathbf{B}\right) - \nabla \times \left(\eta \nabla \times \mathbf{B}\right) ,
\end{equation}
for a prescribed velocity field $\mathbf{v}\left(r,\theta,\phi,t\right)$ and prescribed turbulent diffusivity $\eta\left(r\right)$. There is no small-scale $\alpha$-effect. Equation \ref{induction2} is solved in a spherical shell using a finite volume scheme. For more details see Appendix A of \citet{kd3}.

Unlike previous 2D B-L models, KD3 explicitly models the buoyant emergence of flux tubes through the convection zone \citep{fan09}. In the 2D models, the active region emergence process has either been parametrised through a volumetric $\alpha$-effect term in the induction equation, or through manual deposition of regions at the surface, corresponding to areas of strong toroidal field at the base of the convection zone \citep[e.g.][]{durney97,nandy01,mj10,guerrero12}. The deposition method has also been used in another 3D B-L dynamo model, developed by \citet{miesch3d} \citep[see also][]{miesch16,karakmiesch17,karakmiesch18,hazra17,hazra18}. However, these `non-local' methods make magnetic flux conservation difficult to enforce because the process of forming the emerging region from the pre-existing toroidal field is not followed explicitly through the induction equation (\ref{induction2}).

In KD3, a time-dependent velocity perturbation is included which is intended to capture the effects of advection and buoyancy on the flux tubes. A similar emergence method has been used by \citet{kumar3d} and \citet{kumar19}. The non-axisymmetric perturbation has a radial component, which transports the tube outwards through the convection zone to the surface; a vortical component, which models the helical convective motions and gives rise to tilts in the active regions; and a diverging component, responsible for expanding the tube as the density decreases. The tube centre velocity is set so that the travel time from $r = 0.7\,R_{\odot}$ to $r = R_{\odot}$ is 25 days, after which the perturbation is removed. This method ensures the conservation of magnetic flux during the emergence process. This is in contrast to the deposition method, where active regions are disconnected and an interior structure is assumed only close to the surface.

\begin{figure}
	\resizebox{\hsize}{!}{\includegraphics{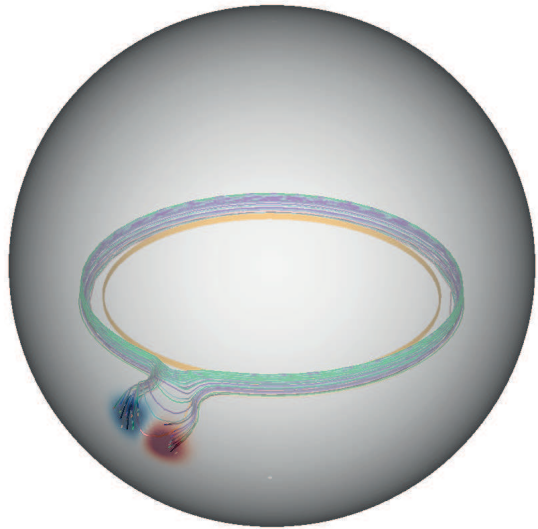}}
	\caption{Three-dimensional image of an emerged active region in KD3. Magnetic field lines are connected to the toroidal field at the base of the convection zone and the radial magnetic field is shown at the transparent surface.}
	\label{region_br}
\end{figure}

Although the KD3 emergence approach is flux-conserving, and \citet{kd3} showed that the model is able to reproduce the qualitative behaviour of active region decay at the surface, leading to polewards transport of flux and reversal of the polar field, closer inspection has shown that the quantitative details of the surface evolution are significantly different from 2D surface flux transport (SFT) models, even when the same horizontal flows and diffusivity and same initial $B_r$ are used at the surface. As an example, the SFT evolution of a single bipolar magnetic region (BMR), shown in Fig. \ref{region_br} and placed at 10\textdegree{} latitude with flux \num{1e22}\,Mx and a tilt angle of 30\textdegree{}, is shown in the top panel of Fig. \ref{bflycomp}, and the KD3 equivalent is shown below. The parameters used are the same as those given in Sect. \ref{sect3}. The BMR is inserted in the SFT simulation at the time when the flux has stopped emerging in KD3, that is, when the unsigned flux at the photosphere has reached its peak (Fig. \ref{fluxcomp}). Even though the differential rotation, meridional flow and horizontal diffusion in the SFT model match the surface parameters of the KD3 simulation, the transport to the poles is faster in the SFT case. In addition, the top panel of Fig. \ref{fluxcomp} shows that there is significantly more flux present at the surface in the KD3 system. There is also a large difference between the respective evolutions of the polar flux (bottom panel of Fig. \ref{fluxcomp}). In KD3, the south polar field barely develops by the end of the simulation, and the peak of the north polar field is stronger and occurs three years later than in the SFT case.\\

\begin{figure}
	\resizebox{\hsize}{!}{\includegraphics{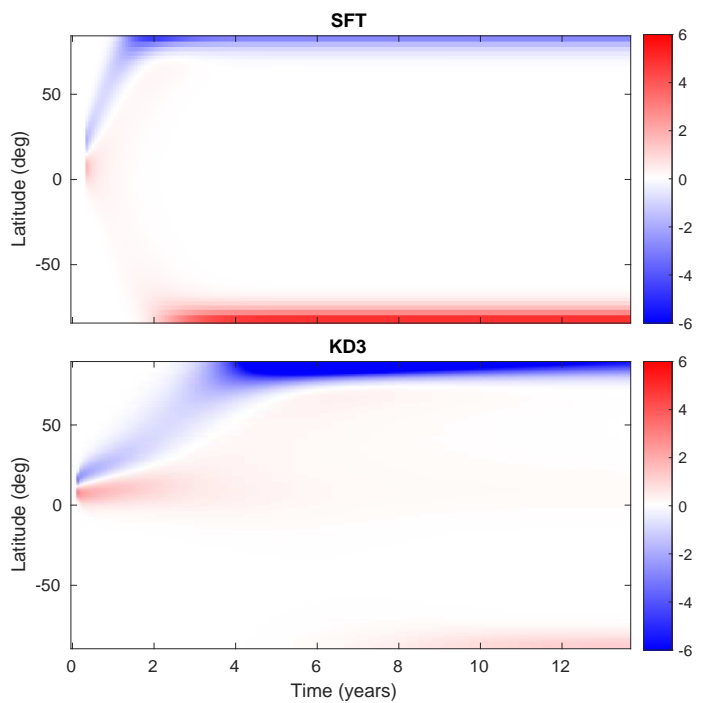}}
	\caption{Top: Longitude-averaged evolution of $B_r$ for a single BMR in a 2D SFT model. Bottom: Surface component of the 3D dynamo model showing the equivalent evolution of the same BMR.}
	\label{bflycomp}
\end{figure}

\begin{figure}
	\resizebox{\hsize}{!}{\includegraphics{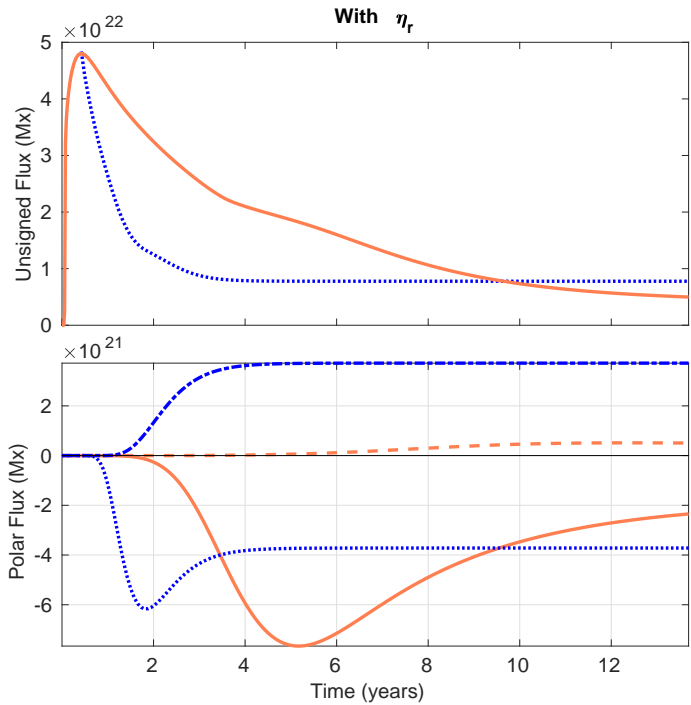}}
	\caption{Top: Comparison of unsigned surface flux from the 2D SFT simulation (blue) and 3D dynamo simulation (orange). Bottom: Comparison of northern (solid and dotted lines) and southern (dashed and dash-dotted) polar flux from the same two simulations, where polar flux is defined as the flux polewards of 70\textdegree{} latitude.}
	\label{fluxcomp}
\end{figure}

Given that the SFT model parameters have been carefully calibrated to match the evolution of $B_r$ on the real Sun \citep{lemerle,me17}, we start from the premise that it is the KD3 model that needs to be modified. In this paper, we show that the incorrect evolution of surface flux in the KD3 model arises from the fact that BMRs remain connected to the base of the convection zone for several years after emergence, owing to the low diffusivity in the convection zone. This effect is illustrated in Sect. \ref{sect2} using a simplified Cartesian 2D model, where we show that increasing the convection zone diffusivity can improve the surface evolution. In Sect. \ref{sect3}, we verify that the same is true in KD3 when simulating a single region. However, Sect. \ref{sect4} shows that increasing the diffusivity in a full-cycle simulation of KD3 has a catastrophic effect on the dynamo. The implications for future 3D modelling are discussed in Sect. \ref{conclusions}.

\section{2D model of active region decay} \label{sect2}

We begin by investigating a 2D model that illustrates the basic cause of the difference between the KD3 and SFT models. Inspired by \citet{avb98}, we take a 2D $\Omega$-loop representing a newly-emerged BMR in the convection zone and evolve it according to diffusion alone. The benefit of a simpler toy model is that it captures the diffusive effects of a 3D model but is computationally less expensive, and at this stage we are not interested in other features such as the amount of poloidal field produced.

\begin{figure}
	\resizebox{\hsize}{!}{\includegraphics{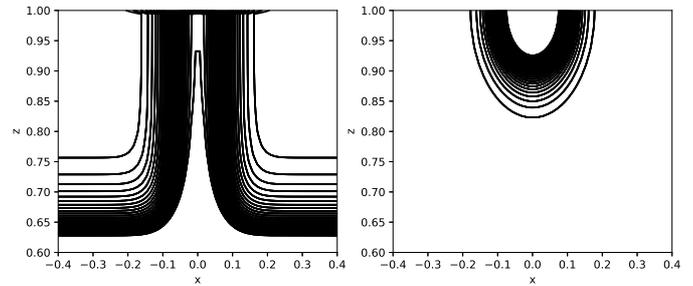}}
	\caption{Two initial conditions used in this paper: an active region connected to the toroidal field (left) and an active region disconnected from the toroidal field (right).}
	\label{init_cond}
\end{figure}

Here we use Cartesian co-ordinates ($x$,$z$) which denote the width and depth of the convection zone domain respectively, with $-0.4 \leq x \leq 0.4$ and $0.6 \leq z \leq 1$. Neglecting variation in the $y$-direction, we write $\mathbf{B}$ in terms of a flux function as:
\begin{equation}
  \mathbf{B} = \nabla\times(A\,\mathbf{e}_y).
\end{equation}
Neglecting advection, Equation \ref{induction2} reduces to
\begin{equation} \label{diffeqn}
  \frac{\partial{A}}{\partial{t}} = \eta\left(z\right)\nabla^2 A .
\end{equation}
Importantly, we allow the diffusivity $\eta$ to be a function of $z$, so that we can investigate the effect of different diffusivity profiles with depth. The effect of advection will be considered in the 3D simulations of Sect. \ref{sect3}. We also simultaneously evolve a 1D surface diffusion model as the analogue of the SFT model. For visualization a potential-field extrapolation is performed in the corona. For our first initial condition, the region is assumed to have emerged and is connected to the toroidal field at the base of the convection zone (see left-hand panel of Fig. \ref{init_cond}), as in KD3, and is of the form:\\
\begin{equation}\label{initialA0}
  A_0 = \exp\left(-\frac{z-0.6}{0.04}\right) + \frac12\exp\left(\frac{(z-1)^2}{0.4} - \frac{x^2}{0.008}\right) .
\end{equation}

We impose periodic boundary conditions in $x$ and set $\partial {A}/\partial{t} = 0$ at the base ($z = 0.6$). At the surface ($z = 1$) we follow \citet{avb98} and \citet{avb07} by setting:
\begin{equation}
  B_{x,\rm cz} = \beta B_{x,\rm cor} ,
\end{equation}
where $B_{x,\rm cz}$ is the horizontal field at the convection zone boundary and $B_{x,\rm cor}$ is the horizontal component of a potential extrapolation into the corona. Then the parameter $\beta$ determines whether the interior field at the photosphere is matched to the potential field in the corona ($\beta = 1$), or whether it is purely radial ($\beta = 0$), which was the original boundary condition in KD3 (and, indeed, in most other models). This will allow us to assess the effect of the top boundary condition on radial diffusion, although for most tests we set $\beta = 0$.

\begin{figure}
	\resizebox{\hsize}{!}{\includegraphics{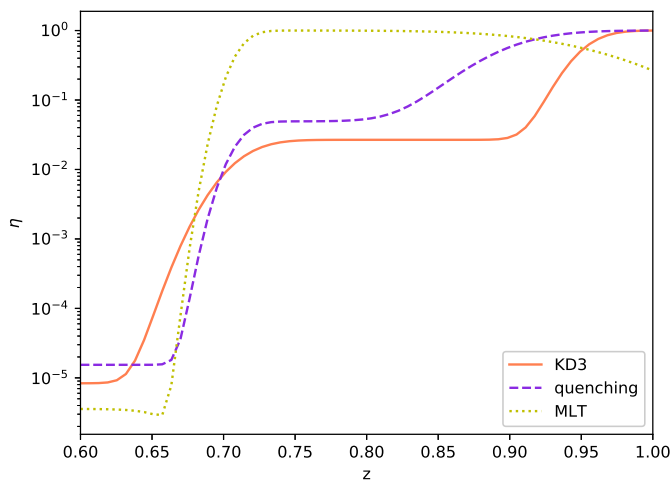}}
	\caption{Normalised multi-step diffusion profiles used in this paper, against a log-scale. The solid orange curve is from KD3, the dashed purple curve is the profile which takes into account diffusivity quenching, and the dotted yellow curve is derived from mixing-length theory.}
	\label{diff_profiles}
\end{figure}

In general we will use the following depth-dependent two-step profile for $\eta \left(z\right)$:
\begin{align}\label{eta_eqn}
  \eta\left(z\right) =& \,\frac{1}{\eta_{\rm max}}\Bigg[\eta _c + \frac{\eta_0 - \eta_c}{2} \left(1 + \mbox{erf}\left(\frac{z - R_1}{\Delta_1}\right)\right)\nonumber\\
  &+ \frac{\eta_s - \eta_0 - \eta_c}{2} \left(1 + \mbox{erf}\left(\frac{z - R_2}{\Delta_2}\right)\right)\Bigg] ,
\end{align}
where $\eta_{\rm max}$ is chosen such that the maximum value of the diffusivity is 1. Here $\eta_c$ is the core diffusivity, $\eta_0$ is the diffusivity in the convection zone, and $\eta_s$ is the surface diffusivity. The step locations and thicknesses are $R_i$ and $\Delta_i$ respectively. The profiles used in this paper are shown in Fig. \ref{diff_profiles}. The solid orange curve shows the profile used by \citet{kd3} in KD3, with parameters $\eta_c = 10^8$\,cm$^2$\,s$^{-1}$, $\eta_0 = \num{1.6e11}$\,cm$^2$\,s$^{-1}$, $\eta_s = \num{6e12}$\,cm$^2$\,s$^{-1}$, $R_1 = 0.71$, $\Delta_1 = 0.03$, $R_2 = 0.95$ and $\Delta_2 = 0.025$. The other profiles will be described later in the section. For a given diffusion profile and boundary condition, Equation \ref{diffeqn} is solved using an explicit finite difference method with Euler timestepping.\\

\begin{figure*}
	\centering
	\resizebox{0.91\hsize}{!}{\includegraphics{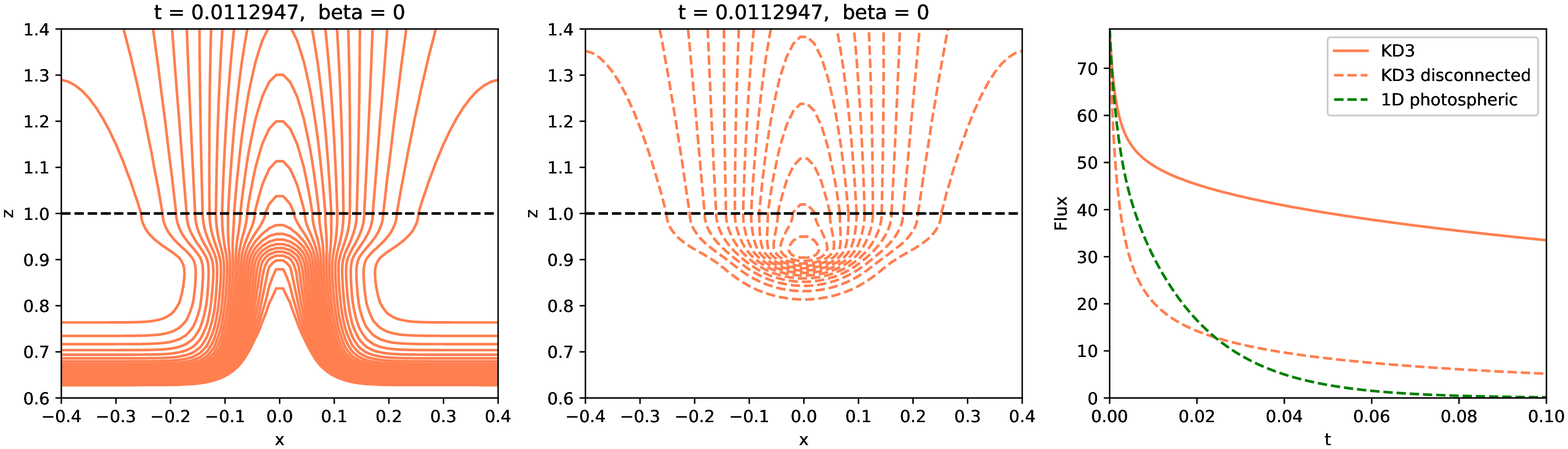}}\\
	\resizebox{0.91\hsize}{!}{\includegraphics{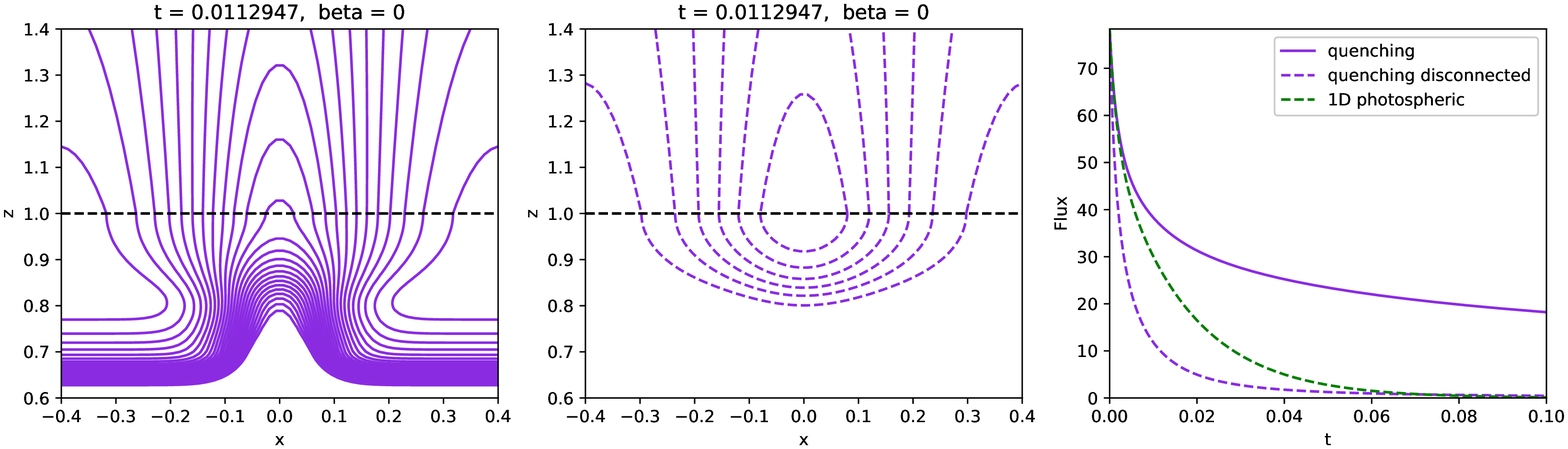}}\\
	\resizebox{0.91\hsize}{!}{\includegraphics{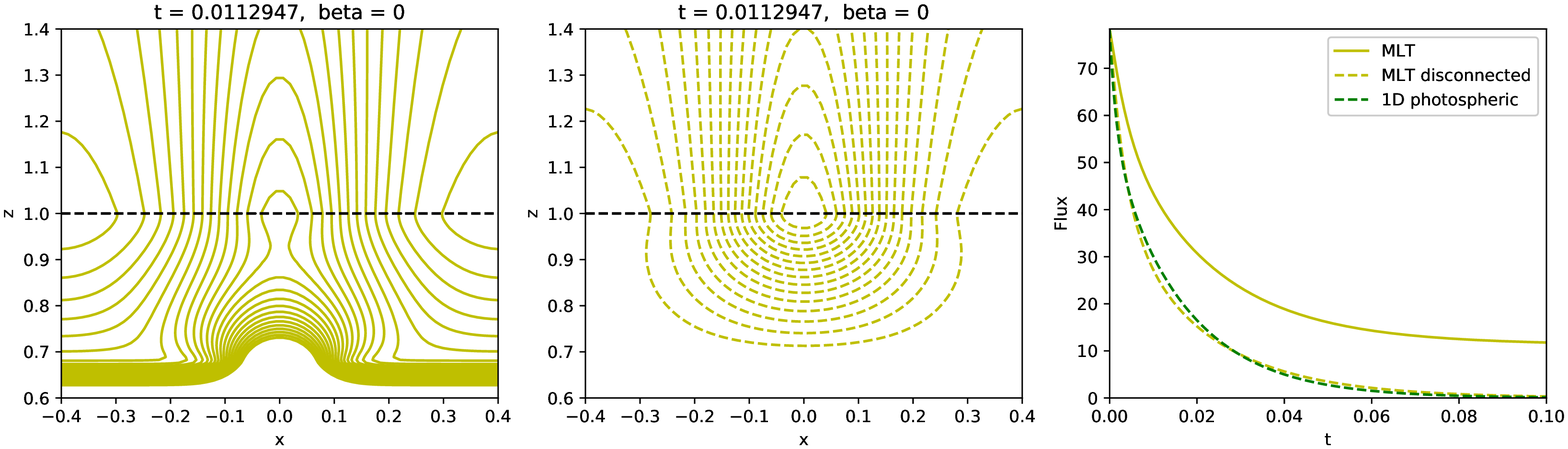}}\\
	\resizebox{0.91\hsize}{!}{\includegraphics{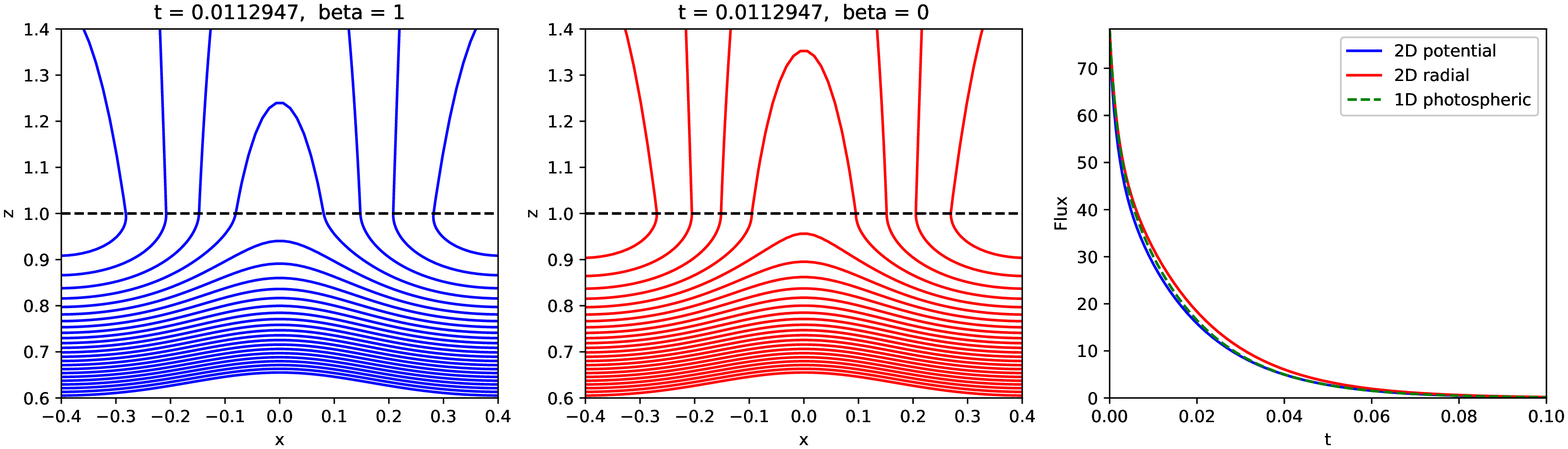}}
	\caption{Snapshots of magnetic field lines from simulations using the KD3 (top row), quenching (second row), MLT (third row) and constant (bottom row) diffusion profiles. For the first three profiles, the left-hand column shows the case where the region is connected to the base, and the middle column shows the simulation with a disconnected initial condition. The black dashed line is the top of the domain, above which is shown a potential-field extrapolation. The right-hand column shows the evolution of magnetic flux at the surface, compared to a 1D surface model (green dashed line). The bottom row (constant $\eta$) compares the effects of potential (blue) and radial (red) boundary conditions but using the same connected initial condition.}
	\label{diff_kd3_hazra}
\end{figure*}

When the KD3 diffusion profile is used, it is clear from the top right panel of Fig. \ref{diff_kd3_hazra} that there is significantly more flux at the surface than would be expected without radial derivatives, as we saw for the KD3 model in Sect. \ref{intro}. This is because the relatively low diffusion below $z=0.9$ does not allow for much diffusive transport, and field lines remain attached to the toroidal field at the base of the convection zone. Because the field lines are fixed in place, movement at the surface is heavily restricted and cancellation at the boundary is limited, resulting in an excess of surface flux. This transpires even though diffusion is stronger near the surface, as indicated by the outwards bulging of field lines.\\

To demonstrate the different evolution for a disconnected active region such as considered by \citet{miesch3d} (hereafter STABLE), we alter the initial condition slightly from Equation \ref{initialA0}:
\begin{equation}\label{initialA02}
A_0 = \frac12\exp\left(-\frac{x^2 + (z-1)^2}{0.008}\right) .
\end{equation}
This forms a potential field below the surface, disconnected completely from the base of the convection zone (see right-hand panel of Fig. \ref{init_cond}).

The top middle panel of Fig. \ref{diff_kd3_hazra} shows a snapshot from the simulation with the original KD3 diffusion profile and disconnected initial condition. Because field lines are no longer connected to the toroidal field at the base of the domain, the weak diffusion no longer plays a role in anchoring field lines in place. This allows for more diffusive transport and cancellation of magnetic flux at the surface. Instead of flux cancellation occurring at the side boundary after field lines are pushed outwards as before, cancellation takes place between the two polarities of the active region. The consequence is that there is less surface flux in the early stages of evolution in comparison to the 1D model. The cancellation rate eventually decreases, but we observe in the top right panel that there is less surface flux present at the end of the simulation than in the case where the connected initial condition was used. The upshot is that the disconnected region qualitatively provides a better match to the surface than the connected region.

In the presence of strong magnetic fields, turbulent diffusivity can be suppressed \citep{roberts75}. This `quenching' can be included in models via a non-linear relationship whereby the diffusion parameter $\eta$ is scaled by the reciprocal of the square of the magnetic field \citep[e.g.][]{tobias96,gilman05,mj08,guerrero09}. By instead taking the geometric spatiotemporal average over many effective diffusivity profiles, \citet{andres11} approximated the effect of the dynamically quenched diffusion using a fixed profile in the form of Equation \ref{eta_eqn} by applying the following parameters: $\eta_c = 10^8$\,cm$^2$\,s$^{-1}$, $\eta_0 = \num{1.6e11}$\,cm$^2$\,s$^{-1}$, $\eta_s = \num{3.25e12}$\,cm$^2$\,s$^{-1}$, $R_1 = 0.71$, $\Delta_1 = 0.017$, $R_2 = 0.895$ and $\Delta_2 = 0.051$. This is shown as the dashed purple curve in Fig. \ref{diff_profiles}, and will henceforth be referred to as the `quenching profile' for simplicity.

A snapshot from the simulation using the quenching profile is displayed in the second row of Fig. \ref{diff_kd3_hazra}. When the original connected initial condition is prescribed, the field lines diffuse downwards initially, but approximately halfway through the simulation the direction of motion changes and the magnetic field starts to diffuse upwards. We note a reduction in the surface flux, presumably because the stronger diffusivity levels extend deeper into the domain and the field lines have more freedom to move, allowing for more diffusive transport. However, we find again that flux cancellation is hindered by the weak diffusion in the lower convection zone, which keeps the field lines attached to the toroidal field.

If instead we start with a disconnected region (middle panel), we find that, as for the KD3 profile, flux cancels inwardly because field lines are not connected to the base of the convection zone. However, it diffuses at a much faster rate than the regime with the KD3 diffusion profile (and hence the 1D case), and by the end of the simulation the majority of the surface flux has been cancelled.

The third profile we experiment with is derived from mixing-length theory \citep[MLT;][]{prandtl25}. \citet{andres11} used the solar interior model of \citet{cdetal} to estimate the mixing-length parameter $\alpha_p$ and hence the diffusivity profile based on GONG data. The value of diffusion found for the convection zone is up to two orders of magnitude larger than those used in KD3 and other kinematic dynamo simulations in literature. This is because simulated dynamo action has not yet been achieved in flux transport dynamos with such strong diffusion. \citet{andres11} attempted to reconcile the MLT estimates with numerical values by incorporating diffusivity quenching, leading to the quenching profile above. Nevertheless, a fit to the MLT profile was also made in the form of Equation \ref{eta_eqn}, with the following resulting parameters: $\eta_c = 10^8$\,cm$^2$\,s$^{-1}$, $\eta_0 = \num{1.4e13}$\,cm$^2$\,s$^{-1}$, $\eta_s = 10^{10}$\,cm$^2$\,s$^{-1}$, $R_1 = 0.71$, $\Delta_1 = 0.015$, $R_2 = 0.96$ and $\Delta_2 = 0.09$. This profile is the dotted yellow curve shown in Fig. \ref{diff_profiles}.

A snapshot from the corresponding simulation is shown in the third row of Fig. \ref{diff_kd3_hazra}. With this diffusion profile and connected initial condition, the field initially diffuses downwards before being pushed back up due to the diffusion gradient at the surface. This surface flux then diffuses to the boundary where it cancels. Low diffusivity at the base means the field still remains attached to the toroidal field but a much larger diffusivity throughout the convection zone helps transport flux upwards from as deep as $z = 0.7$. We see that field lines are being pushed together at the top of the domain due to the reduced diffusivity near the surface and a balance between outwards and inwards diffusion. At a higher cadence, we observe that this causes the field lines to reconnect. The position of the null initially moves downwards, before changing direction and reaching the surface after approximately a third of the simulation time. After this point, magnetic field diffuses outwards rapidly. In terms of surface flux, this regime is closer to the 1D case than any other two-step profile we test with the connected active region, though still not a good match.

The middle panel shows field lines from the simulation with the MLT profile and STABLE initial condition. Because of the strong diffusivity in the bulk of the domain, the field spreads out in the convection zone and diffuses radially outwards due to the reduced diffusivity at the surface. This leads to a surface evolution that matches the 1D case very closely.

We now assess the effect of the upper boundary condition on the surface evolution. For this test, we prescribe a constant diffusivity of $\eta = 1$ independent of depth. A snapshot of the simulation is shown in the bottom row of Fig. \ref{diff_kd3_hazra}. The left-hand panel shows magnetic field lines where the upper boundary condition is potential, and the middle panel shows the field lines where the boundary condition is radial. Qualitative differences are small, but we see in the right-hand panel that there is a little too much magnetic flux at the surface in the radial case, compared to the 1D surface model. Conversely, the potential case matches the 1D evolution closely. If $\beta = 1$, we introduce $-\partial B_x/\partial z$ into Equation \ref{diffeqn} at the surface which is not present in the radial case. Hence the difference between the two regimes is only situated in the upper quarter of the domain. The enforcement of a radial field at the surface boundary also means that the field lines interact with the periodic boundary later because they are strictly vertical, as opposed to the potential case where cancellation can occur more readily. Since the diffusivity is high throughout the domain, field lines can move freely, and the majority of the flux is diffused out of the convection zone by the time we reach $t = 0.1$.

Although the choice of radial or potential-field boundary condition can slightly change the amount of magnetic flux at the surface, the differences are only small, and starker differences arise when we prescribe a more realistic multi-step diffusion profile in place of the constant diffusivity, or change the connectivity of the active region. Further tests show that the small improvement attained by changing boundary condition is the same regardless of the choice of diffusion profile. Interestingly, the 2D model in the constant case provides a good match to the 1D model and explains in part why the MLT profile performs best out of the multi-step profiles we tested: the strong diffusivity allows the magnetic field to diffuse outwards in both cases, the only difference being that the field lines remain attached to the toroidal field in the MLT case due to a weak base diffusion.

The periodic boundary conditions in $x$ can be interpreted as the presence of neighbouring active regions. To check the influence of this inter-region spacing, we tried increasing the width of the domain. This results in more flux present at the surface because it takes longer to diffuse to the boundary and cancel. However, the results above hold qualitatively, and in any case we cannot choose the locations of active region emergence when simulating the evolution of observed BMRs, so varying the width of the domain does not give us significantly deeper insight.

\section{Effect of diffusivity for a 3D decaying active region} \label{sect3}

We return to the 3D dynamo model KD3 to test whether the results found in Sect. \ref{sect2} hold qualitatively here as well. We emerge a single region at 10\textdegree{} latitude with flux \num{1e22}\,Mx and a tilt angle of 30\textdegree{}. Once the region has emerged after 25 days, the velocity perturbation is turned off. A snapshot of the system is taken on that day, and all subsequent experiments are run from time of emergence. This best reflects the scenario modelled in the simplified 2D diffusion model in Sect. \ref{sect2}, and is the same setup as described in Figs. \ref{region_br}--\ref{fluxcomp} but starting from a different time.

Differential rotation takes the form of \citet{char99}:
\begin{align}
  \Omega\left(r,\theta\right) =&\, \Omega_c + \frac12\left[1 + \mbox{erf}\left(\frac{r - R_0}{\Delta_0}\right)\right]\Big[\Omega_E - \Omega_c \nonumber\\
  & + \left(\Omega_p - \Omega_E\right)\left(C\cos^2\theta + \left(1-C\right)\cos^4\theta\right)\Big] ,
\end{align}
where $\Omega_c = \num{2.71434e-6}$\,s$^{-1}$, $\Omega_E = \num{2.9531e-6}$\,s$^{-1}$, $\Omega_p = \num{2.07345e-6}$\,s$^{-1}$, $C = 0.483$, $R_0 = 0.7\,R_\odot$ and $\Delta_0 = 0.025\,R_\odot$.

For meridional flow we first define the following stream function \citep{kd3}:
\begin{align}
  \Psi\left(r,\theta\right) =& \frac{-v_0 \left(r - R_p\right)}{7.633\,r\sin\theta} \sin\left(\pi\frac{r - R_p}{R_\odot - R_p}\right) \nonumber\\
  & \cdot \exp\left[-\left(\frac{r - R_1}{\Gamma}\right)^2\right]\Big[1 - \exp\left(-1.5\theta^2\right)\Big] \nonumber\\
  & \cdot \left\{1 - \exp\left[1.8\left(\theta - \frac{\pi}{2}\right)\right]\right\} ,
\end{align}
where $R_p = 0.62\,R_\odot$, $R_1 = 0.1125\,R_\odot$, $\Gamma = \num{3.47e8}$\,m and $v_0 = 20$\,m\,s$^{-1}$. Then the meridional circulation is given by
\begin{equation}
  \mathbf{v_m} = \frac{1}{\rho\left(r\right)} \nabla \times \big[\Psi\left(r,\theta\right)\mathbf{e_{\phi}}\big] ,
\end{equation}
where $\rho \left(r\right) = \left(\dfrac{R_\odot}{r} - 0.95\right)^\frac32$ is the radial density profile.

\begin{figure}
	\centering
	\resizebox{\hsize}{!}{\includegraphics{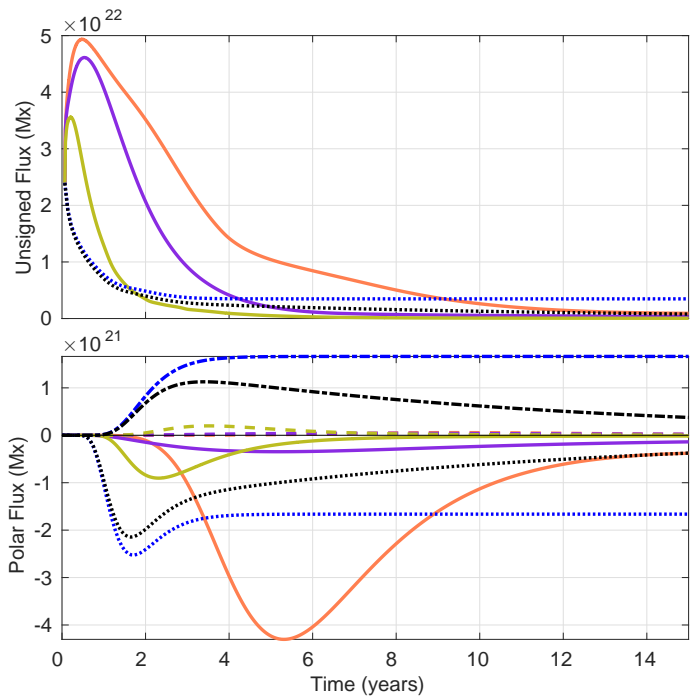}}
	\caption{Top: Unsigned surface flux from 3D simulations of a single active region, using the KD3 diffusion profile (orange), quenching profile (purple) and MLT profile (yellow). The equivalent 2D SFT flux is shown in the dotted blue (without decay) and black (with decay) curves. Bottom: northern polar flux (solid and dotted curves) and southern polar flux (dashed and dash-dotted curves) from the same simulations.}
	\label{diag_nodr}
\end{figure}

We use a grid resolution of $\Delta_r = \frac{0.45\,R_\odot}{48}$ in radius, and a variable grid in latitude and longitude (see \citealp{kd3} for details). Here we set the equatorial grid spacing $\Delta_{\phi} = \frac{2\pi}{384}$. Initial and boundary conditions are the same as used by \cite{kd3}: the bottom boundary condition at $r = 0.55\,R_\odot$ is a perfectly conducting core, meaning $\partial{\left(r B_{\theta}\right)}/\partial{r} = \partial{\left(r B_{\phi}\right)}/\partial{r} = 0$. The upper boundary condition is radial, although we expect from Sect. \ref{sect2} that changing to a potential-field boundary condition would have a negligible effect on the flux evolution. The initial condition is created by emerging a single BMR from a layer of toroidal field at the base of the convection zone of the form:
\begin{equation}\label{inittor}
  \mathbf{B} = -\frac{B_0}{2}\left(\frac{\cos\theta}{|\cos\theta|}\right)\left[\mbox{erf}\left(\frac{r - R_7}{\Delta_8}\right) - \mbox{erf}\left(\frac{r - R_8}{\Delta_8}\right)\right] \mathbf{e_{\phi}} ,
\end{equation}
with $R_7 = 0.66\,R_\odot$, $R_8 = 0.74\,R_\odot$, $\Delta_8 = 0.018\,R_\odot$ and $B_0 = \num{2.5e3}$\,G. At the surface, we do not prescribe any initial magnetic field, aside from that of the emerged BMR. Diffusivity is now given by $\eta_s\,\eta(r/R_\odot)$ using Equation \ref{eta_eqn}, where $\eta_s = \num{6e12}$\,cm$^2$\,s$^{-1}$. We run the model using a different diffusion profile from Sect. \ref{sect2} each time. The resulting unsigned surface flux and polar flux profiles are shown in Fig. \ref{diag_nodr}.\\

As we found in the 2D model (Sect. \ref{sect2}), the KD3 profile (orange) restricts cancellation, due to the weak diffusivity keeping field lines connected to the toroidal field. This results in a vast excess of flux at the surface. Qualitatively, the other profiles also exhibit the same behaviour as in the 2D model. The MLT profile (yellow) provides a more rapid decay of flux due to the increased diffusivity in the convection zone allowing for disconnection, and the quenching profile (purple) lies somewhere between the other two. By the end of the simulation, there is a similar amount of surface flux in the KD3 simulations as in the SFT simulations, as shown by the dotted blue and black curves. These curves represent the models without the exponential decay term \citep[see][]{me17}, and with a decay parameter of $\tau = 10$ years, respectively. Whilst the decay term in the SFT model makes only a very small difference in the total unsigned surface flux, its impact at the poles is more evident, acting as a sink for the polar flux which is not otherwise possible in the SFT model \citep{baumann06}. Although the peak strength of the northern polar field is weaker in the MLT case than the SFT model, it occurs at a similar time and the shape of the profile is close to that of the SFT model when exponential decay is included. In summary, these experiments with the decay of a single active region suggest that increasing the diffusivity in the bulk of the convection zone can improve the realism of long-term surface flux evolution compared to the original KD3 model.

\section{Effect of diffusivity on a 3D full-cycle simulation} \label{sect4}

\begin{figure}
	\resizebox{\hsize}{!}{\includegraphics{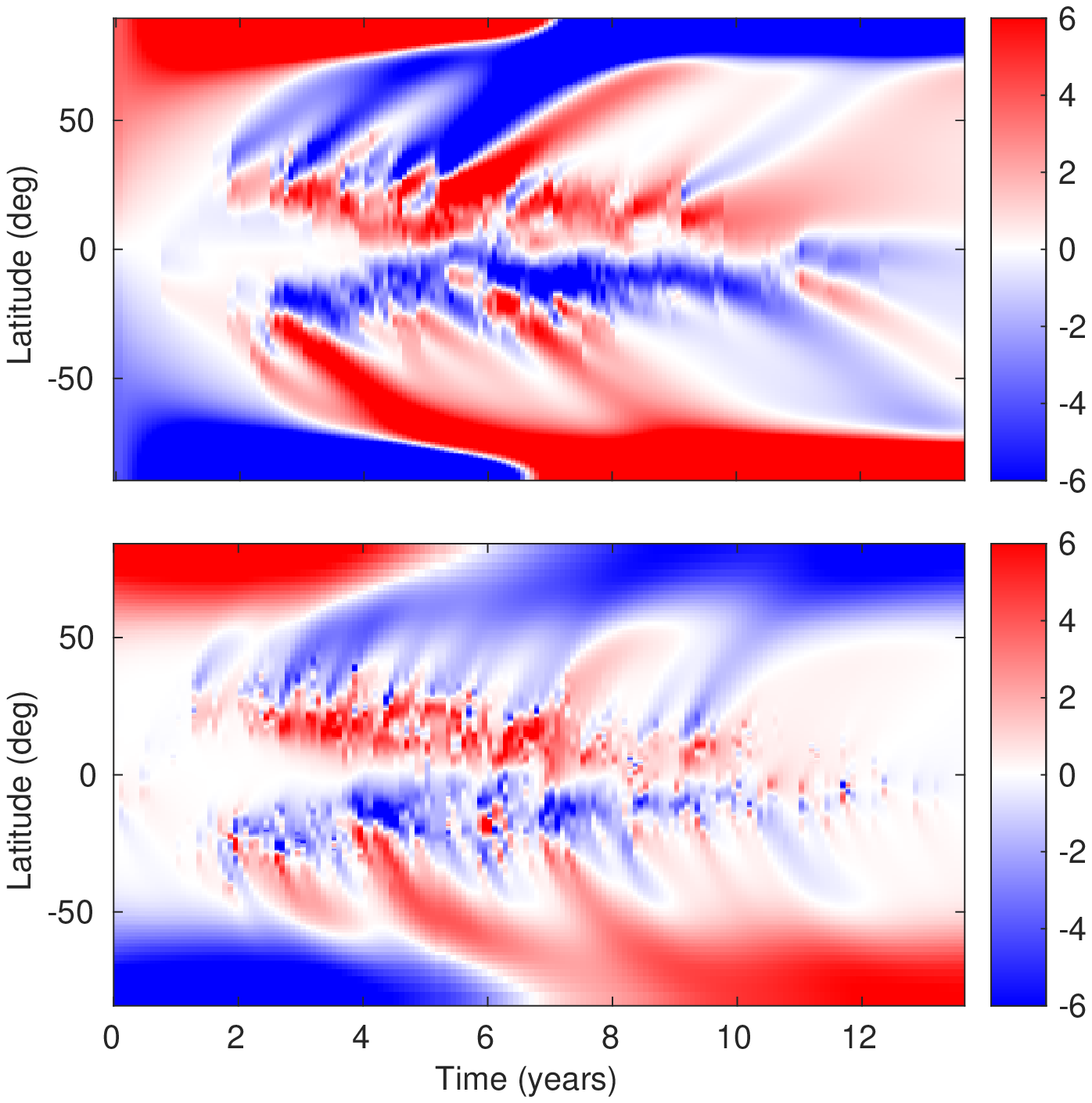}}
	\caption{Top: Simulation of Cycle 23 from \citet{kd3}. Bottom: Optimal butterfly diagram of Cycle 23 from \citet{me17}.}
	\label{fullcycle_kd3}
\end{figure}

\citet{kd3} demonstrated a simulation of a full solar cycle using BMR data from Solar Cycle 23. However, this was not systematically calibrated to observations. It can be seen in the top panel of Fig. \ref{fullcycle_kd3} (or equivalently Fig. 12 of \citealp{kd3}) that the magnetic field is too strong and poleward surges are too slow compared to the optimal butterfly diagram found by \citet{me17}, shown in the lower panel of Fig. \ref{fullcycle_kd3}, which was calibrated against observations. The active regions across the full solar cycle behave similarly to the individual region in Fig. \ref{bflycomp}. We repeat this 3D simulation of Cycle 23 but replace the original diffusion profile (orange curve in Fig. \ref{diff_profiles}) with the quenching and mixing-length theory profiles (purple and yellow curves in Fig. \ref{diff_profiles} respectively).\\

Equation \ref{inittor} again defines the initial toroidal field, but now we try $B_0 = 250$\,G. An initial dipolar field is given by
\begin{equation}
  \mathbf{B} = \nabla \times \left(A_{\phi} \mathbf{e_{\phi}}\right) ,
\end{equation}
where
\begin{equation}
  A_{\phi} = B_d\,\frac{\sin\theta}{r^3}\left(\frac{r - 0.7\,R_\odot}{0.3\,R_\odot}\right) ,
\end{equation}
and $A_{\phi} = 0$ for $r < 0.7\,R_\odot$ \citep{jouve08}. The field strength is set as $B_d = -0.008\,B_0$.

\begin{figure}
	\resizebox{\hsize}{!}{\includegraphics{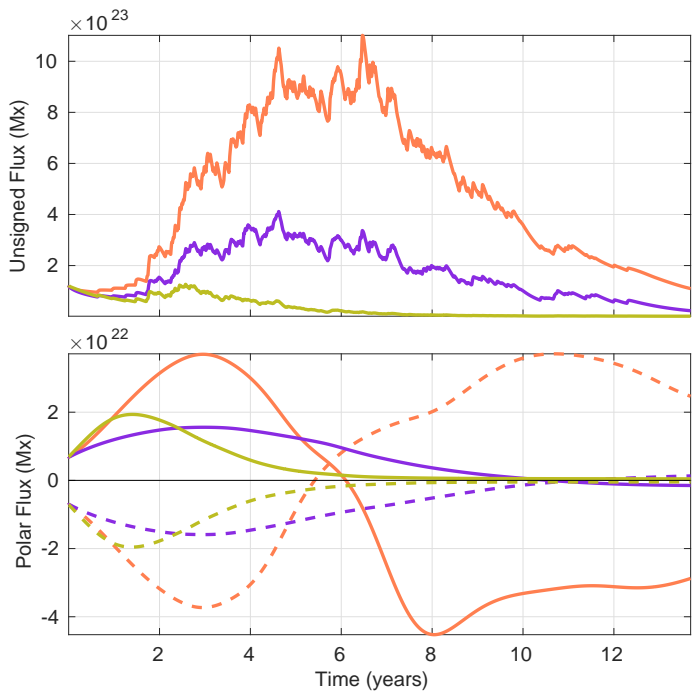}}
	\caption{Top: Unsigned surface flux from 3D simulations of Cycle 23 using the KD3 diffusion profile (orange), quenching profile (purple) and MLT profile (yellow). Bottom: northern polar flux (solid line) and southern polar flux (dashed line) from the same simulations.}
	\label{diag_fullcycle}
\end{figure}

We run the simulation for 5000 days, using observed BMRs of Cycle 23 from NSO Kitt Peak as input data \citep{yeates07}, as in \citet{kd3}. The unsigned surface flux and signed polar flux for the simulation of \citet{kd3} are shown by the orange curves (top and bottom respectively) in Fig. \ref{diag_fullcycle}. The purple profiles in this plot correspond to the simulation where the quenching diffusivity profile has been used. If all parameters other than the diffusivity profile are fixed, it is evident that not enough magnetic flux reaches the surface, and the polar field is barely able to reverse.\\

\begin{figure}
	\resizebox{\hsize}{!}{\includegraphics{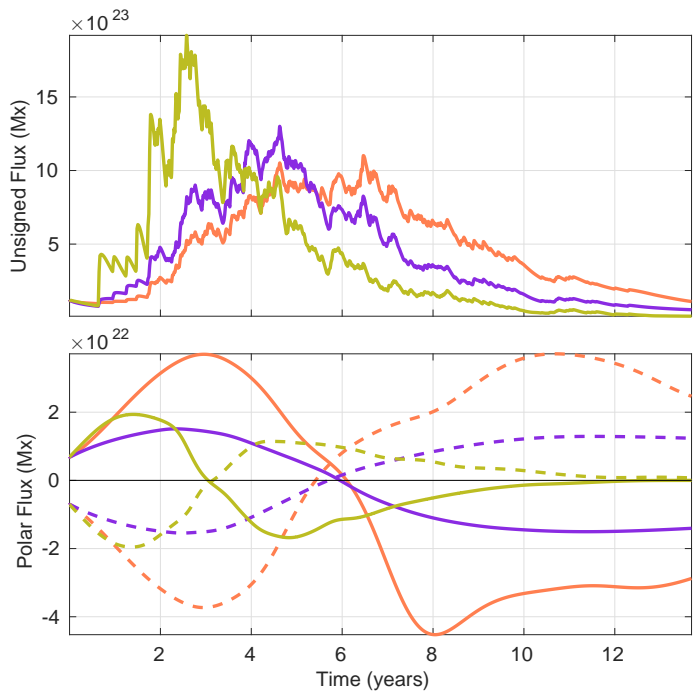}}
	\caption{Top: Unsigned surface flux from 3D simulations of Cycle 23 using the KD3 diffusion profile (orange), quenching profile (purple) and MLT profile (yellow), but where the initial toroidal field has been strengthened by one and two orders of magnitude for the latter two respectively. Bottom: northern polar flux (solid line) and southern polar flux (dashed line) from the same simulations.}
	\label{diag_fullcycle_beq10}
\end{figure}

\begin{figure}
	\resizebox{\hsize}{!}{\includegraphics{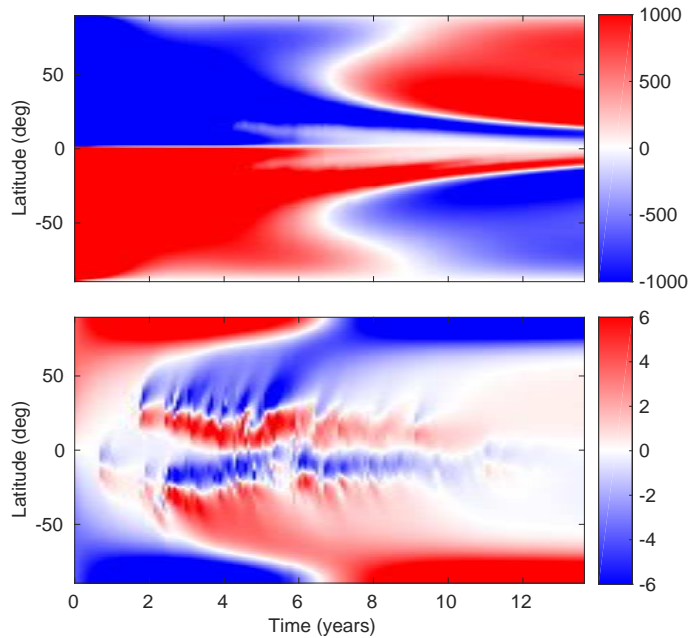}}
	\caption{Top: Toroidal field at the base of the convection zone from a 3D simulation of Cycle 23 using the quenching profile and a strengthened initial toroidal field. Bottom: Radial magnetic field at the surface from the same simulation.}
	\label{bfly_fullcycle_quench_beq10}
\end{figure}

To combat this, we increase the strength of the initial toroidal field by an order of magnitude. This provides a stronger source from which active regions can develop, thereby increasing the amount of flux at the photosphere. This is demonstrated by the purple curve in the top panel of Fig. \ref{diag_fullcycle_beq10}. Here, the total surface flux peaks earlier than the original simulation. In the bottom panel, we see that the polar field reverses at a similar time to the original case, albeit with a reduced strength throughout the simulation. Nevertheless, the toroidal field appears to be strong enough to produce more regions as a subsequent cycle (top panel of Fig. \ref{bfly_fullcycle_quench_beq10}) if we were to continue the simulation. The bottom panel of Fig. \ref{bfly_fullcycle_quench_beq10} shows the surface butterfly diagram of the same simulation. While it is suboptimal, it displays observable features of the solar cycle and a more realistic distribution and transport of magnetic flux than before. A future task is to calibrate other parameters in the model against observations while keeping the quenching profile fixed.\\

Ideally we would like to be able to simulate Cycle 23 using the diffusion profile derived from mixing-length theory, because the enhanced diffusivity acts as a means of disconnecting the emerged regions from the toroidal field, and this profile gave the closest match to the surface-only evolution in Sects. \ref{sect2} and \ref{sect3}. Figure \ref{diag_fullcycle} shows that even less flux emerges at the surface in this case, because the diffusion in the convection zone is too strong and kills off the majority of rising flux tubes and returning poloidal flux. Even when the initial toroidal field is increased by an order of magnitude, it rapidly diffuses and so no regions are able to emerge after a few years.

\begin{figure}
	\resizebox{\hsize}{!}{\includegraphics{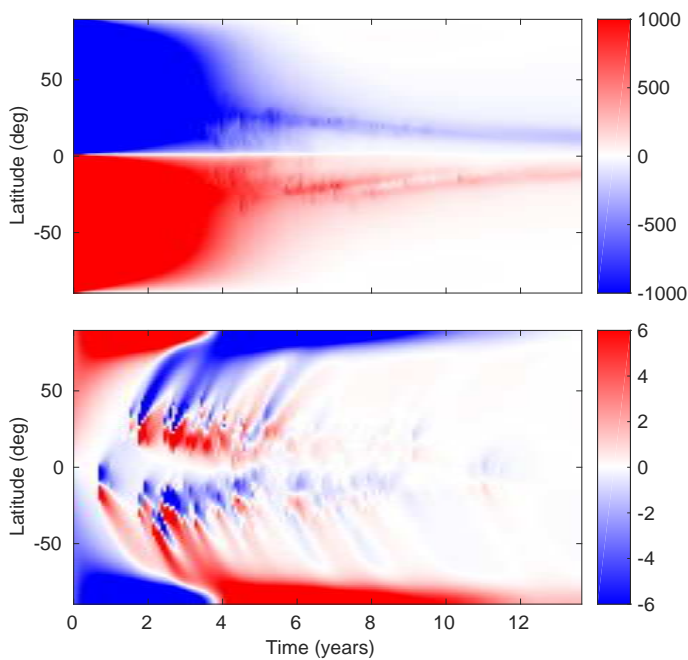}}
	\caption{Top: Toroidal field at the base of the convection zone from a 3D simulation of Cycle 23 using the MLT profile and a strengthened initial toroidal field. Bottom: Radial magnetic field at the surface from the same simulation.}
	\label{bfly_fullcycle_mlt_beq100}
\end{figure}

When the toroidal field is increased by another order of magnitude, the flux still decays too rapidly, as shown by the yellow curve in Fig. \ref{diag_fullcycle_beq10}. However, we now observe polar field reversal, although very early in the cycle, and the bottom panel of Fig. \ref{bfly_fullcycle_mlt_beq100} shows that the surface evolution during the first few years of the cycle appears to be sun-like. The top panel of Fig. \ref{bfly_fullcycle_mlt_beq100} shows that no new toroidal field is created. This occurs in all simulations when the MLT diffusivity profile is used and is one reason why dynamos have thus far been unable to accommodate the diffusion profile derived from mixing-length theory.\\

Scaling the MLT profile by a factor of 0.5 allows significantly more flux to emerge at the surface, but it is still not enough on its own to sustain the dynamo. However, if we also shift the location of the low-diffusivity step in the MLT profile up so that the toroidal field is stored in a region of low diffusion (i.e. set $R_1 = 0.74$ and $\Delta_1 = 0.024$), we find that the field survives for longer and more flux can reach the surface. However, although more new toroidal field starts to appear at the base of the convection zone for the next cycle, it is still too weak, and the polar field at the surface still reverses too early. In summary, increasing the diffusivity to the level required for a realistic surface evolution is not on its own sustainable in a full-cycle simulation, because the high diffusivity removes too much flux from the system.

\section{Conclusions} \label{conclusions}

Our main conclusion is that 3D kinematic dynamo models cannot produce a realistic evolution of magnetic flux on the solar surface if active regions are allowed to remain connected to the base of the convection zone. Within the framework of the KD3 model, where active regions are formed self-consistently through imposed velocity perturbations, the only way to achieve disconnection is through enhanced turbulent diffusivity in the convection zone. Whilst such an enhanced diffusivity is compatible with estimates from mixing-length theory, flux transport dynamo models have been unable to function with such a high diffusivity \citep{andres11}. In this paper we have demonstrated that, indeed, a full-cycle simulation with KD3 is not possible with such strong diffusivity, despite the fact that it leads to a realistic surface flux evolution when simulating the decay of a single active region.

\begin{figure}
	\begin{minipage}{.5\linewidth}
		\centering
		\includegraphics[width=\linewidth]{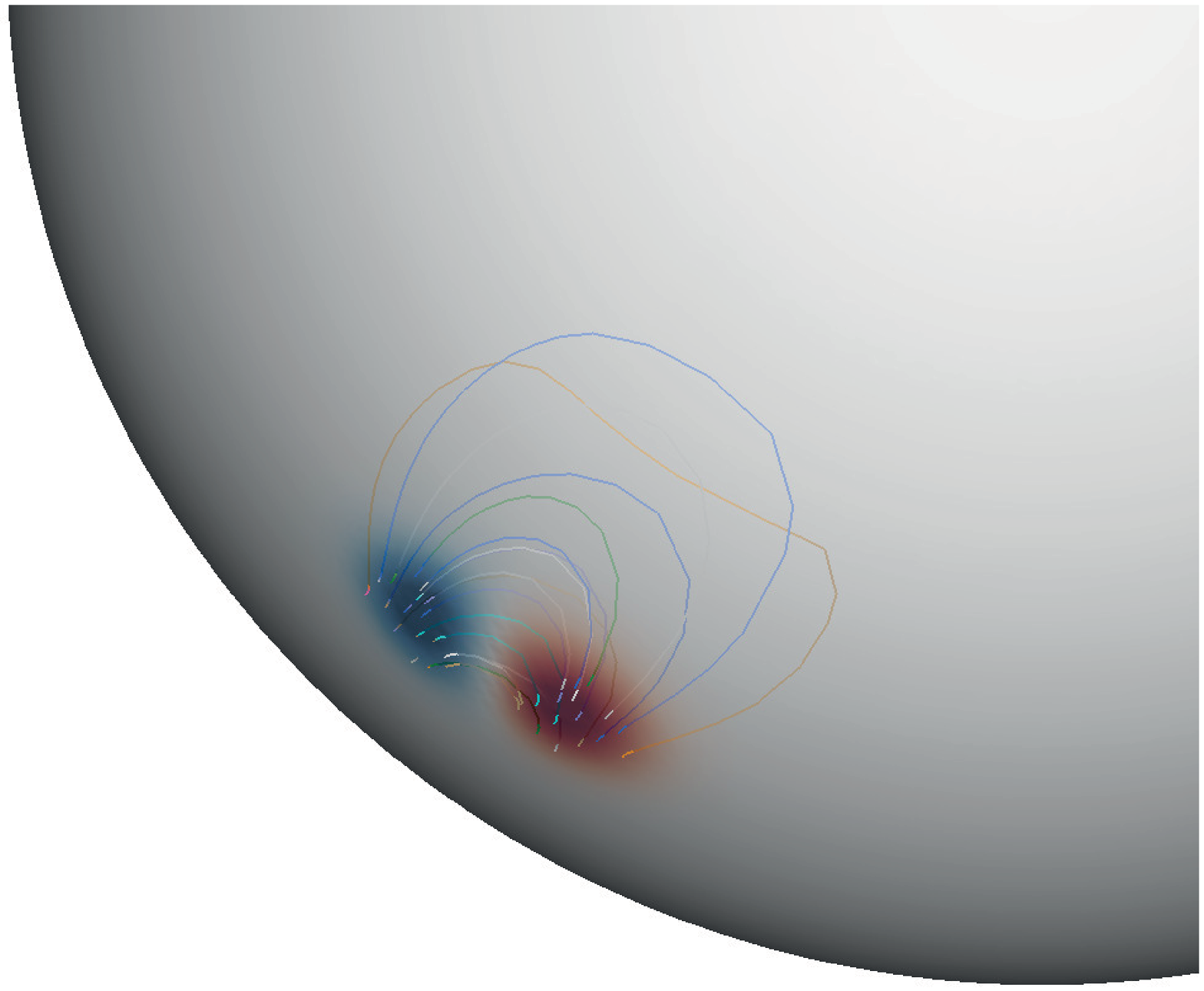}
	\end{minipage}%
	\begin{minipage}{.5\linewidth}
		\centering
		\includegraphics[width=\linewidth]{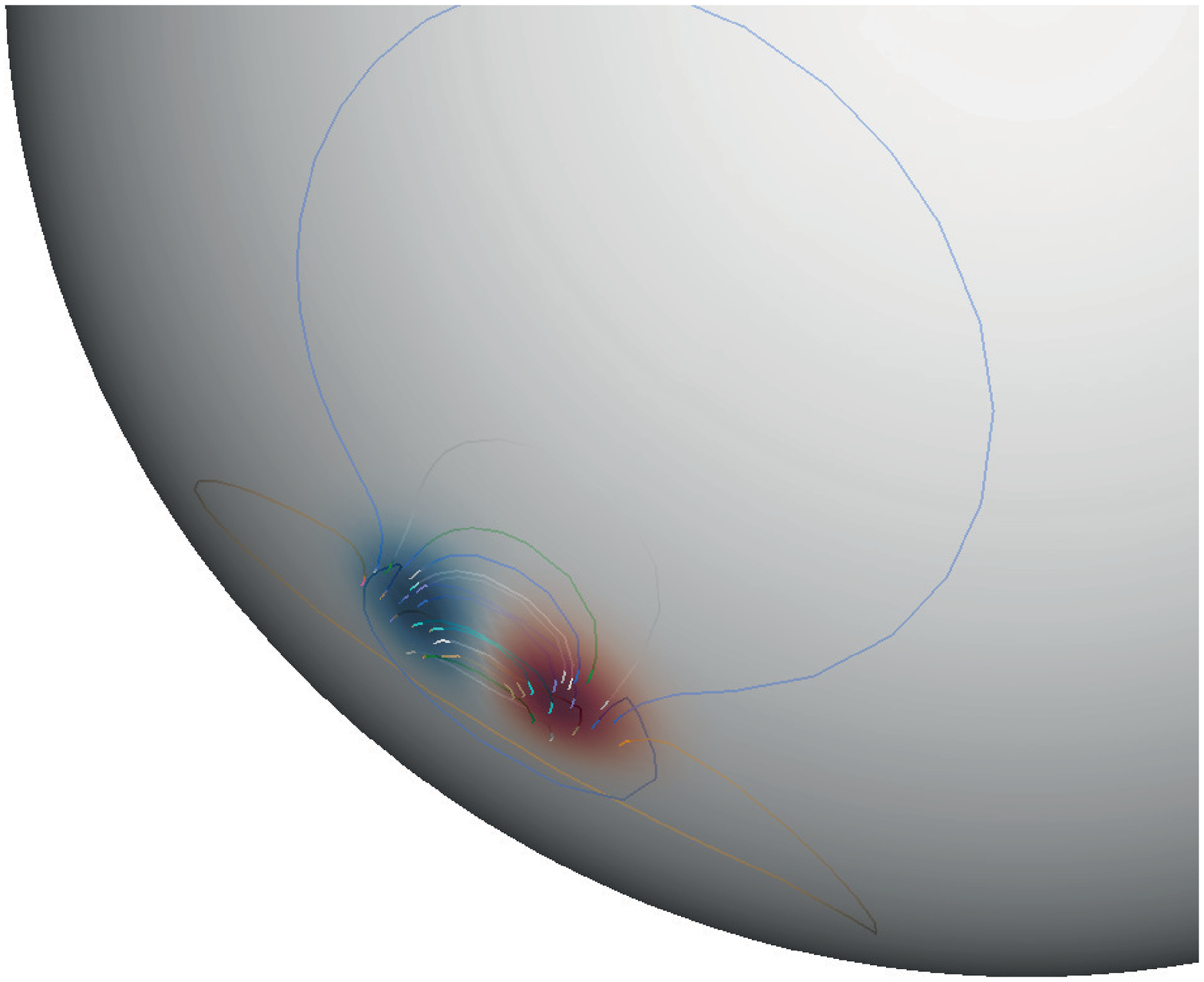}
	\end{minipage}
	\begin{minipage}{\linewidth}
	    \centering
	    \includegraphics[width=\linewidth]{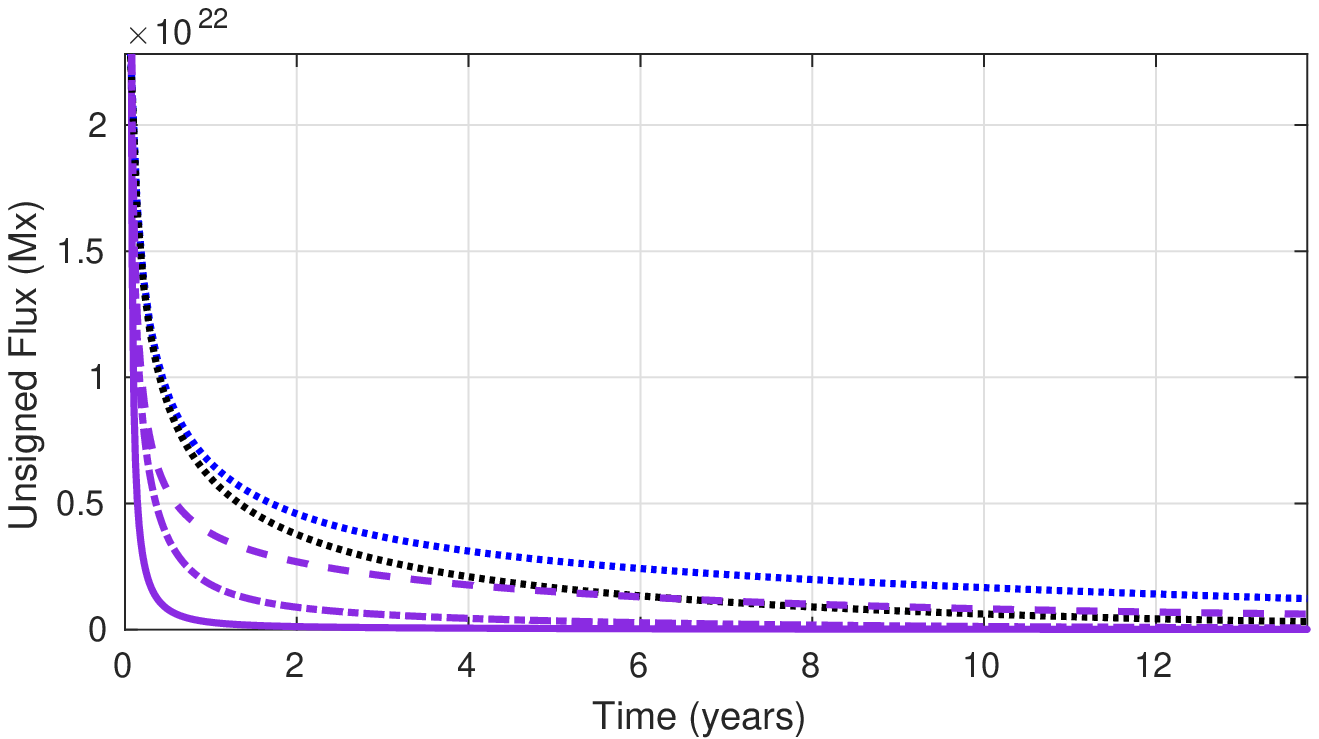}
    \end{minipage}
    \caption{Top: Three-dimensional images of a disconnected active region in KD3 with depth $0.65\,R_\odot$ (left) and $0.95\,R_\odot$ (right). The radial magnetic field is shown at the transparent surface and field lines below the surface. Bottom: Unsigned surface flux from 3D simulations of a single disconnected active region with depth $0.65\,R_\odot$ (dashed), $0.8\,R_\odot$ (dash-dotted) and $0.95\,R_\odot$ (solid), using the quenching profile. The equivalent 2D SFT flux is shown in the dotted blue (without decay) and black (with decay) curves.}
	\label{3dplots}
\end{figure}

A possible resolution to this problem is suggested by simulations where the active region is initially disconnected from the base of the convection zone, as in the 3D dynamo model STABLE of \citet{miesch3d}. With magnetic field lines no longer anchored to the base of the convection zone, diffusion is much more effective. We have shown clearly in our 2D simplified model (Sect. \ref{sect2}) that this leads to a better fit with the surface evolution, for a given diffusivity profile. In the bottom panel of Fig. \ref{3dplots}, we demonstrate the effect of disconnecting regions in KD3, using the quenching diffusion profile as an illustration (purple curves). For the interior magnetic field we calculate a potential field extrapolation inward from the given surface $B_r$ with a perfectly conducting boundary condition at a fixed depth. We find that the flux decays faster than in the SFT case, but that the evolution is dependent on the depth of the potential field. If the region is shallow (solid curve), it decays very quickly, but if we increase the depth to $0.8\,R_\odot$ (dash-dotted curve) and then $0.65\,R_\odot$ (dashed curve), we observe an increasingly better fit to the 2D surface evolution, although it is still by no means perfect. With further study, it may be possible to produce an excellent fit to the surface model using disconnected regions. The top row of Fig. \ref{3dplots} shows two examples of disconnected regions from KD3 with depths $0.65\,R_\odot$ (left) and $0.95\,R_\odot$ (right). The shallower depth forces the outermost field lines to be pressed close to the surface (cf. Fig. \ref{region_br}).\\

Several observational facts at the surface point to the likelihood that active regions become rapidly disconnected from their roots after emergence, on a timescale of days \citep{fan09}. For example, if they remained connected we might expect to see a relaxation of tilt angle towards the east-west line once emergence (and the resulting Coriolis force) ceased to operate, as well as a continued separation of the two polarities reflecting conditions at the roots of the emerged flux tube. Indeed, \citet{fan94} remarked that the success of SFT models in reproducing the surface evolution is itself evidence that active regions are no longer dynamically constrained from below. The actual process of disconnection is less understood, although \citet{schrijver99} argued that subsurface reconnection on the required timescale will be a natural consequence of convective motions. \citet{fan94} proposed a mechanism of `dynamical disconnection', based on the idea that the magnetic field in the subsurface legs of the active region could be weakened to sub-equipartition values as it tries to establish hydrostatic equilibrium, thus enabling reconnection. This was confirmed in a 1D calculation by \citet{schussremp05}.

Let us end with some broader remarks. In this paper, we have treated the SFT model as the `ground truth,' but in reality we must remember that 2D SFT models have their own limitations. The resulting errors are likely too small to change our broad conclusions in this paper, but may be non-negligible. For example, Figure \ref{diag_nodr} shows how adding an exponential decay term to the SFT model improves the qualitative match with the 3D model when viewed over solar-cycle timescales. This term is a crude parametrization of radial diffusion, which is missing in the basic SFT model \citep{baumann06} but included consistently in the 3D model. Our results suggest that, for the diffusivities used in typical flux-transport dynamo simulations, this term will have a non-negligible effect on the surface evolution.

At this point, the important question remains as to whether a flux-conserving 3D B-L model like KD3 could match the observed surface flux evolution on the Sun. From our results, we expect that the existing dynamo codes with non-local active region deposition, like STABLE, are better able to match the surface evolution than the existing KD3 code. However, KD3 nevertheless has the advantage of satisfying the magnetic induction equation during the emergence of active regions, allowing the critical flux budget to be correctly accounted for. In future, it would therefore be desirable to develop an active region emergence model that satisfies the induction equation while at the same time achieving rapid disconnection of active regions, perhaps through controlled reconnection.  Exploring the full parameter space of such a model, including the meridional flow and turbulent pumping profiles, would be a significant computational task. Nevertheless, the work in this paper, along with restrictions on the surface parameters from \citet{me17}, does add some valuable constraints to this optimization problem.

\begin{acknowledgements}

TW thanks Durham University Department of Mathematical Sciences for funding his PhD studentship. ARY thanks STFC for financial support through grant ST/N000781/1. This research was supported by the NASA Living With a Star Grant NNH15ZDA001N.

\end{acknowledgements}

\bibliographystyle{aa}
\bibliography{mybiblio}

\end{document}